\documentclass[useAMS,usegraphicx]{mn2e}
\newcommand\beq{\begin{equation}}
\newcommand\eeq{\end{equation}}

\newcommand\reffig[1]{Fig.~\ref{fig:#1}}

\title[Lensing of Anisotropic Sources]
{Gravitational Lensing of Anisotropic Sources}

\author[Perna \& Keeton]
{Rosalba Perna$^1$ \& Charles R. Keeton$^2$ \\
$^1$JILA and Department of Astrophysical
and Planetary Sciences, University of Colorado, Boulder, CO 80309, USA \\
$^2$Department of Physics and Astronomy, Rutgers University,
136 Frelinghuysen Road, Piscataway, NJ 08854, USA
}

\begin{document}

\date{Draft \today}

\maketitle

\begin{abstract}
In strong gravitational lensing, the multiple images we see correspond
to light rays that leave the source in slightly different directions.
If the source emission is anisotropic, the images may differ from
conventional lensing predictions (which assume isotropy).  To identify
scales on which source anisotropy may be important, we study the angle
$\delta$ between the light rays emerging from the source, for
different lensing configurations.  If the lens has a power law profile
$M\propto R^\gamma$, the angle $\delta$ initially increases with lens
redshift and then either diverges (for a steep profile $\gamma<1$),
remains constant (for an isothermal profile $\gamma=1$), or vanishes
(for a shallow profile $\gamma>1$) as $z_{\rm l} \to z_{\rm s}$.  The
scaling with lens mass is roughly $\delta \propto M^{1/(2-\gamma)}$.
The results for an NFW profile are qualitatively similar to those for
a shallow power law, with $\delta$ peaking at about half the redshift
of the source (not half the distance).  In practice, beaming could
modify the statistics of beamed sources lensed by massive clusters:
for an opening angle $\theta_{\rm jet}$, there is a probability as
high as $P \sim 0.02$--$0.07\,(\theta_{\rm jet}/{0.5}^\circ)^{-1}$
that one of the lensed images may be missed (for $2 \la z_{\rm s} \la
6$).  Differential absorption within Active Galactic Nuclei could
modify the flux ratios of AGNs lensed by clusters; a
sample of AGNs lensed by clusters could provide further constraints on
the sizes of absorbing regions.  Source anisotropy is not likely to be
a significant effect in galaxy-scale strong lensing.
\end{abstract}

\begin{keywords}
cosmology: gravitational lensing --- quasars: absorption lines ---
galaxies: jets --- galaxies: active -- gamma rays: bursts
\end{keywords}

\section{Introduction}

In gravitational lensing studies, the emission from the background
source is usually assumed to be isotropic such that the appearance
of lensed images depends only on the mass distribution in the
foreground lensing object and the angular position of the source
with respect to the lens.  The fluxes of the images can then be
written as $F_i = \mu_i\,F_{\rm src}$ where $\mu_i$ is the lensing
magnification at the position of image $i$, and the source flux
$F_{\rm src}$ is assumed to be the same for all images.  In this
case we can interpret observed flux ratios simply as lensing
magnification ratios: $F_j/F_i = \mu_j/\mu_i$.  (The one caveat is
that if the source is variable, differences in light travel times
make it necessary to monitor lensed images and synchronize the
light curves in the source time frame before taking the ratio;
e.g., Eigenbrod et al.  2005; Kochanek et al. 2006; Fohlmeister
et al. 2007, 2008.)  The ability to associate flux ratios with
magnification ratios underlies many lensing applications, including
using anomalous flux ratios to constrain dark matter substructure
(e.g., Metcalf \& Madau 2001; Chiba 2002; Dalal \& Kochanek 2002;
Keeton, Gaudi \& Petters 2003, 2005; Chiba et al. 2005), and using
lens statistics to constrain the mass function and density profiles
of galaxies and clusters (e.g., Keeton \& Madau 2001; Kochanek \&
White 2001; Takahashi \& Chiba 2001; Oguri 2002; Ma 2003; Kuhlen,
Keeton, \& Madau 2004; Oguri \& Keeton 2004; Oguri \& Blandford 2009)
\footnote{In lens statistics, the image fluxes
are used to determine whether multiple images will be detectable
and hence whether systems will be identified as lenses.}

However, many astrophysical sources relevant for gravitational
lensing have some degree of anisotropy in their emission.  One
example is provided by Gamma-Ray Bursters (GRBs).\footnote{While
there are currently no confirmed cases of GRB lensing, the possibility
has received considerable attention (e.g., Paczynski 1986; Mao 1992;
Grossman \& Nowak 1994; Holz et al. 1999; Nemiroff et al. 2000;
Porciani \& Madau 2001).} Numerical simulations (e.g., MacFadyen
\& Woosley 1999) indicate that the local emissivity is a strong
function of the angle that the line of sight makes with the jet
axis, and the interpretation of afterglow observations appears to
confirm this scenario (Perna, Sari \& Frail 2003). Other
interpretations instead suggest that the emission is concentrated
in a jet with sharp edges and a range of opening angles that can
be as small as a degree scale (Nakar, Granot \& Guetta 2004).
Recent numerical simulations of axisymmetric, magnetically driven
outflows (Komissarov et al. 2009) have shown that the $\gamma$-ray
emitting components of GRB outflows magnetically accelerated are
very narrow, with $\theta_{\rm jet}\la 1^\circ$.  Furthermore, the
GRB emission is highly relativistic, with Lorentz factors
$\Gamma \sim 100$--$300$ (e.g. Piran 2000). This implies that each point
on the emitting surface is only visible to observers within an angle
$\theta_{\rm view} \sim 1/\Gamma \sim 0.2$--$0.5$ deg.

Anisotropy in both emission and absorption is also typical of Active
Galactic Nuclei (AGNs). The fast TeV variability of the blazars Mrk
501 and PKS 2155-304 was interpreted by Giannios, Uzdensky \& Begelman
(2009) as the result of compact emitting regions moving with Lorentz
factors of $\Gamma \sim 100$ embedded within a jet moving at lower
speed.  Nair, Jin \& Garrett (2005) proposed a helical jet
model to explain observations of the gravitationally lensed blazar
PKS 1830$-$211.  Even if the emission is intrinsically isotropic,
absorption by clouds within the broad line absorption (BAL) region
can introduce anisotropy into the net flux out of the source.
Indeed, Chelouche (2003) argued that differential absorption
along multiple lens sightlines could be detected in spectra of
lensed BAL quasars, and Green (2006) suggested that this effect
might explain variability in the broad emission lines of only one
image of the wide-separation lens SDSS J1004+4112 (Richards et al.
2004). The presence of small-scale structure within the AGN
outflow is also supported by numerical simulations (Proga, Stone
\& Kallman 2000).  While the qualitative picture of the BAL region
is generally accepted, the quantitative details are still rather
uncertain.  Debate over the location of BAL clouds spans some
five orders of magnitude, $\sim\!0.01$--$1000$ pc (Elvis 2000;
de Kool et al. 2001; Everett, Konigl \& Arav 2002).  Very little
is known about the size of the clouds, although a number of studies
suggest an upper limit to the size of $\la 10^{14}$ cm (Baldwin et
al. 1995; Elvis 2000; also N. Arav 2009, private communication).
A cloud of this size at a distance of a few parcsecs would produce
differential absorption on a scale of arcseconds.

The key question, then, is whether anisotropy in real astrophysical
sources is likely to have a significant impact on observed strong
lensing.  In order to answer this question, we need to quantify the
angle $\delta$ between light rays as they leave the source on their
way to becoming the multiple images we observe; only anisotropy on a
scale $\la\delta$ will be relevant for lensing.  In \S\ref{sec:geom}
we set up the general problem of lensing of an anisotropic source
by a spherical mass distribution.  In \S\ref{sec:PL} we study how
the angular beam separation $\delta$ depends on the lensing
configuration for a wide range of lenses with power law mass
distributions.  In \S\ref{sec:NFW} we focus on lenses with the
Navarro, Frenk \& White (1996) profile.  In \S\ref{sec:MC} we use
Monte Carlo simulations to compute the distribution of $\delta$
angles for a population of NFW clusters.  We discuss the
implications of our results in \S\ref{sec:summary}.  We adopt the
cosmological parameters $H_0=73$ Mpc$^{-1}$ km s$^{-1}$, 
$\Omega_m=0.24$, and $\Omega_{\Lambda}=0.76$ (Spergel et al. 2007).

\section{Light beam separation in multiply imaged sources}
\label{sec:geom}

\begin{figure}
\begin{center}
\includegraphics[width=0.45\textwidth]{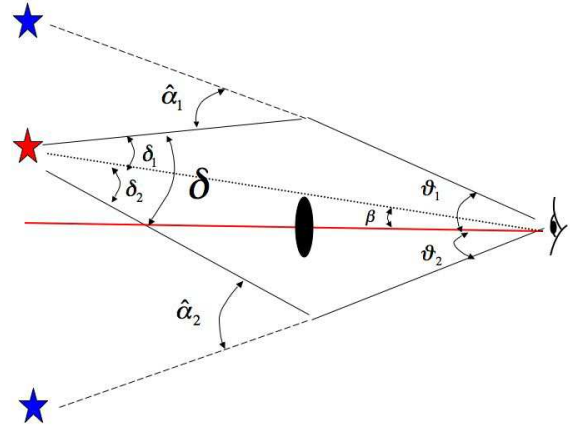}
\end{center}
\caption{
Schematic representation of lensing with two images.  A mass profile
shallower than isothermal would also have a third image that passes
near the galaxy on the opposite side from the source, but such central
images are faint and difficult to detect (e.g., Winn et al. 2004) so
we focus on the two main, outer images.
}
\label{fig:geom}
\end{figure}

\reffig{geom} shows a schematic representation of the gravitational
lensing geometry with two images, which is generic when the source
is sufficiently well aligned with a spherical mass distribution.
When the central density profile is shallower than isothermal, the
image on the opposite side of the lens from the source would be
accompanied by a third image that lies closer to the lens galaxy
(Burke 1981), but such images are rarely observed (e.g., Winn,
Rusin \& Kochanek 2004) so in our study we refer to the two outermost
images.

As shown in the figure, $\beta$ is the angular position of the
(unlensed) source on the sky with respect to the optical axis (the
line connecting the observer and the lens), $\theta_{1,2}$ are the
angular positions of the two main images, and the corresponding
deflection angles are $\hat{\alpha}_{1,2}$.  We are interested in
the angle $\delta$ between the two light rays that emerge from the
source to produce the observed images.  This angle can be written
as the sum of the angles $\delta_{1,2}$ between the light rays and
the line of sight from the observer to the (unlensed) source.  Simple
geometry allows us to identify these angles as
\beq
\delta_1 = \left|\hat{\alpha}_1\right| -\left|\theta_1\right|+\beta\;,
\quad\mbox{and}\quad
\delta_2 = \left|\hat{\alpha}_2\right| -\left|\theta_2\right|-\beta\;,
\label{eq:del12}
\eeq
so we can express the angular beam separation as
\beq
\delta = \left|\hat{\alpha}_1\right| + \left|\hat{\alpha}_2\right| - 
(\left|\theta_1\right|+ \left|\theta_2\right|)\;.
\label{eq:delta}
\eeq
For a spherically symmetric mass distribution, the deflection angle
is
\beq
\left|\hat{\alpha}(\theta)\right| = \frac{4 G M(\theta)}
{c^2 D_{\rm l} \left|\theta\right|}\;,
\eeq
where $M(\theta)$ is the projected mass enclosed within angle $\theta$,
and $D_{\rm l}$ is the angular diameter distance from the observer to
the lens, and the sign of the deflection angle is chosen to match the
sign of $\theta$.  The source and image positions and the deflection
angle are linked via the lens equation,
\beq
\beta = \theta - \frac{D_{\rm ls}}{D_{\rm s}}\,\hat{\alpha}(\theta)\;,
\label{eq:radlens}
\eeq
where $D_{\rm s}$ and $D_{\rm ls}$ are angular diameter distances from
the observer to the source and from the lens to the source, respectively.

\section{Power law mass distribution}
\label{sec:PL}

\subsection{The lens model}
\label{sec:PL1}

To develop a general understanding of how the angular beam separation
depends on the lensing geometry and the physical properties of the
lens, we begin with a simple power law mass profile.  In length units
we write $M(R) = A\,R^\gamma$ with $A$ some constant, so in angular
units we have $M(\theta) = A(D_{\rm l}\theta)^\gamma$.  The cases
$\gamma=0$ and $\gamma=1$ correspond to the familiar cases of a point
mass lens (PM) and a singular isothermal sphere (SIS), respectively.
The lens equation takes the form
\beq
\beta\,=\,\theta\, \mp \,\theta_E^{2-\gamma}\,\left|\theta\right|^{\gamma-1}\;,
\label{eq:powlens}
\eeq
where we use the minus sign when $\theta>0$ and the plus sign when
$\theta<0$, and the angular Einstein radius is
\beq
\theta_E = \left(\frac{4 G A}{c^2} \frac{D_{\rm ls}}
{D_{\rm s}D_{\rm l}^{1-\gamma}}\right)^{\frac{1}{2-\gamma}}\;.
\label{eq:powE}
\eeq
Note that if we consider the mass within some fixed physical radius
we have $M \propto A$ and hence $\theta_E \propto M^{1/(2-\gamma)}$.

For $0\le\gamma<1$ the lens equation (\ref{eq:powlens}) formally has
two solutions for all source positions, although when $\beta$ gets
large the image on the opposite side of the lens is faint.  For
$\gamma=1$ the lens equation has one or two solutions depending on
the position of the source, while for $1<\gamma<2$ it has one or
three solutions.  For each solution, the corresponding deflection
angle is given by
\beq
\left|\hat{\alpha}(\theta)\right| = \frac{D_{\rm s}}{D_{\rm ls}}\;
\theta_E^{2-\gamma}\;\left|\theta\right|^{\gamma-1}\;.
\label{eq:alphapow}
\eeq

\subsection{The angular beam separation}
\label{sec:PL2}

As we study how the angular beam separation depends on the lens
redshift $z_{\rm l}$, we want to keep the {\em physical} properties
of the lens fixed, which is why we elected to write the mass profile
as $M = A\,R^\gamma$ where $A$ is a constant.  To facilitate the
comparison of models with different power law slopes $\gamma$, we
choose the value of $A$ such that the different masses all have the same
Einstein radius when the lens is halfway between the observer and
source.

\begin{figure*}
\includegraphics[width=0.45\textwidth]{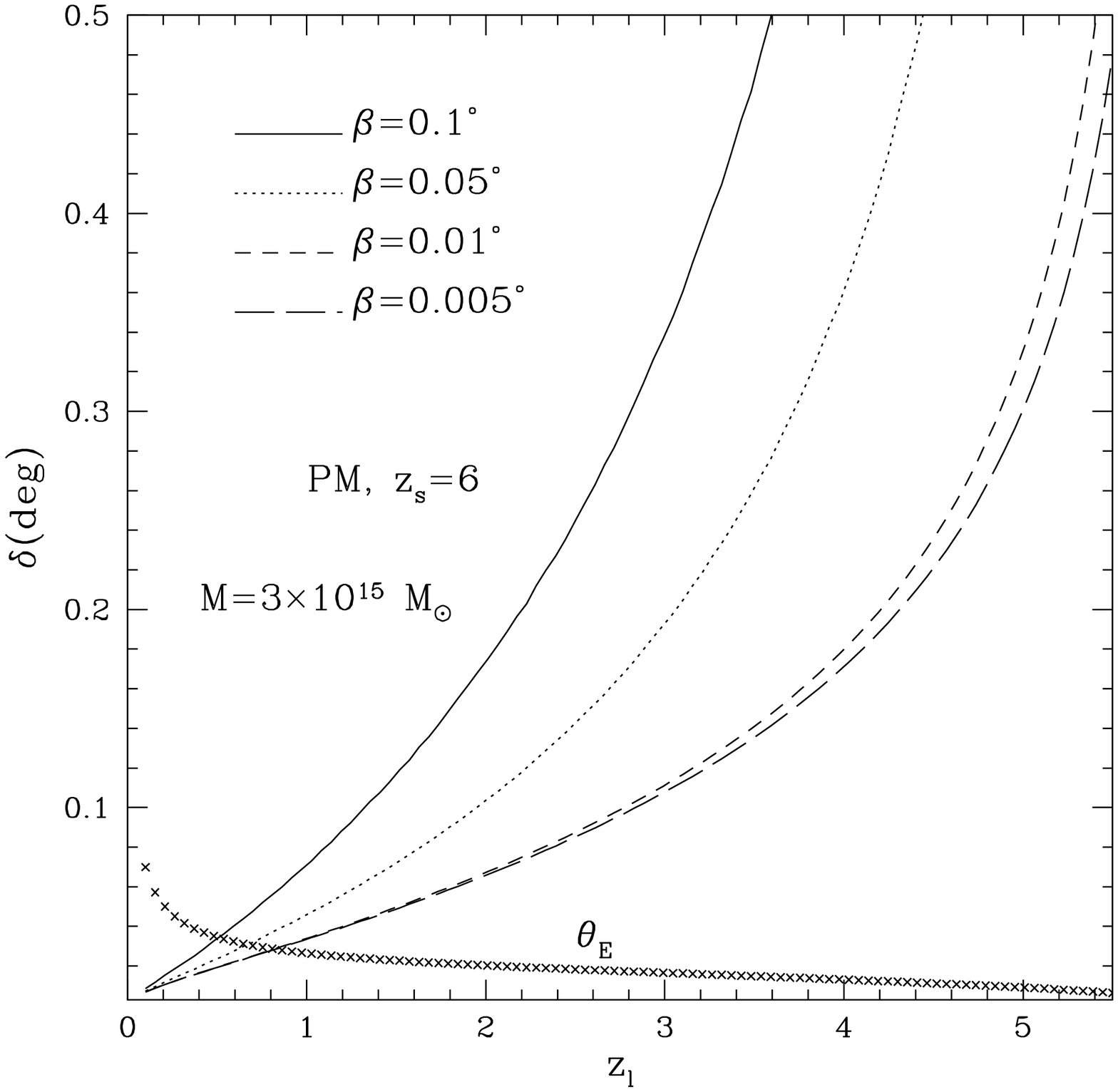}
\includegraphics[width=0.45\textwidth]{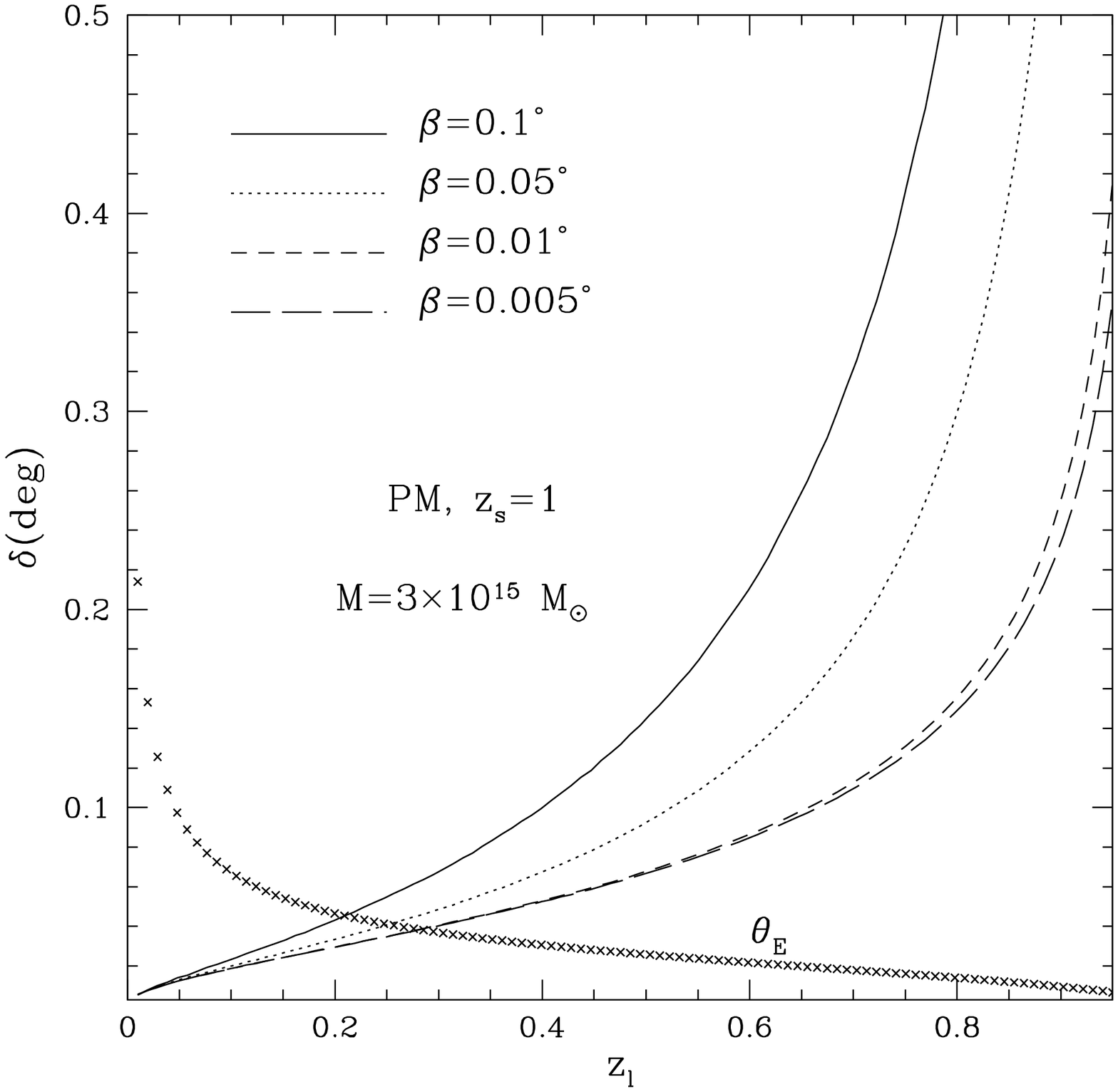}
\caption{
The angular beam separation $\delta$ for a point mass lens as a
function of the lens redshift.  Different curves correspond to
different source positions.  The lens mass is
$M=3\times 10^{15} M_\odot$ and the source redshift is $z_s=6$
{\em (left panel)} and $z_s=1$ {\em (right panel)}.  The points
near the bottom of each panel show the Einstein radius as a function
of the lens redshift.
}
\label{fig:PM1}
\end{figure*}

\reffig{PM1} shows the angular beam separation as a function of lens
redshift for a point mass lens ($\gamma=0$), considering two values
of the source redshift and different values of the angle $\beta$ of
the source with respect to the optical axis.  A striking result is
the steep increase of $\delta$ with $z_l$.  This and other scalings
can be understood as follows.  The two images are located at the
angular positions $\theta_{1,2}=(\beta \pm \sqrt{4\theta_E^2+\beta^2})/2$,
and the corresponding deflection angles are
$\hat{\alpha}_{1,2}=(D_{\rm s}/D_{\rm ls})\;\theta_E^2/\theta_{1,2}$.
Equation (\ref{eq:delta}) then yields
\beq
\delta_{\rm PM}=\left(\frac{D_{\rm s}}{D_{\rm ls}}-1 \right)\,
\sqrt{4\theta_E^2+\beta^2}\;.
\label{eq:delPM}
\eeq
This equation elucidates the trends apparent in the figure.
First, since $\theta_E^{\rm PM}\propto D_{\rm ls}^{1/2}$ we see
that $\delta_{\rm PM}$ formally diverges\footnote{In practice the
small-angle approximation would break down before $\delta$ truly
diverges.} as the lens approaches the source ($z_{\rm l} \to z_{\rm s}$
and hence $D_{\rm ls} \to 0$).  This divergence occurs for all
values of the source angle $\beta$.  Second, it is clear that
$\delta_{\rm PM} \to 0$ as the lens approaches the observer
($z_{\rm l} \to 0$ and $D_{\rm ls} \to D_{\rm s}$).  Third, when
the source and lens redshifts and the lens mass are all fixed,
$\delta_{\rm PM}$ increases with $\beta$.  Fourth, when the source
is well aligned with the lens ($\beta \ll \theta_E$), the angular
beam separation scales with the lens mass as
$\delta_{\rm PM} \propto M^{1/2}$.  Finally, in the opposite limit
in which $\beta \gg \theta_E$, $\delta_{\rm PM}$ becomes independent
of the mass of the lens (although this particular case is perhaps
less relevant than the others because when $\beta \gg \theta_E$
the counter-image is faint).

\begin{figure*}
\includegraphics[width=0.45\textwidth]{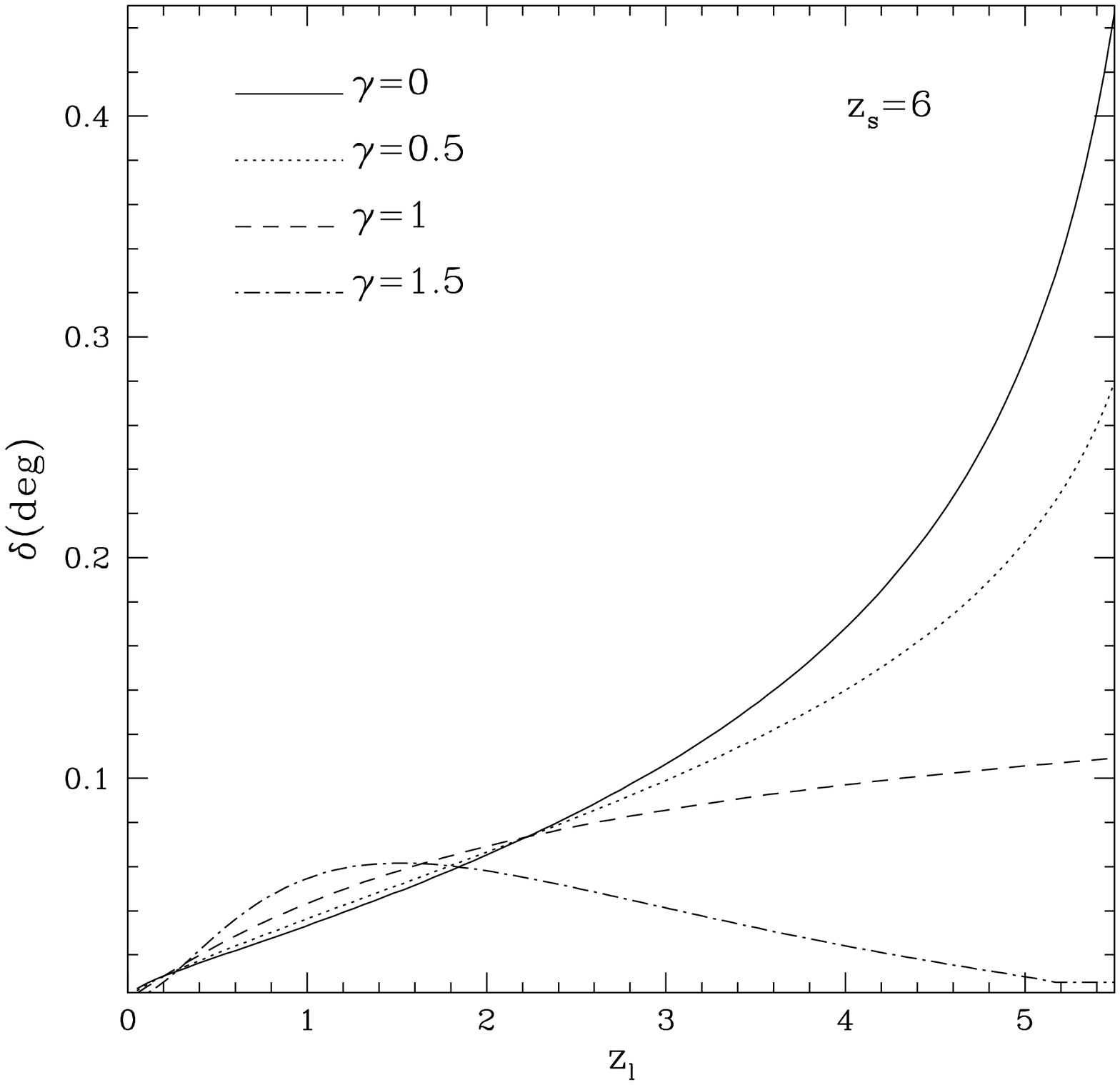}
\includegraphics[width=0.45\textwidth]{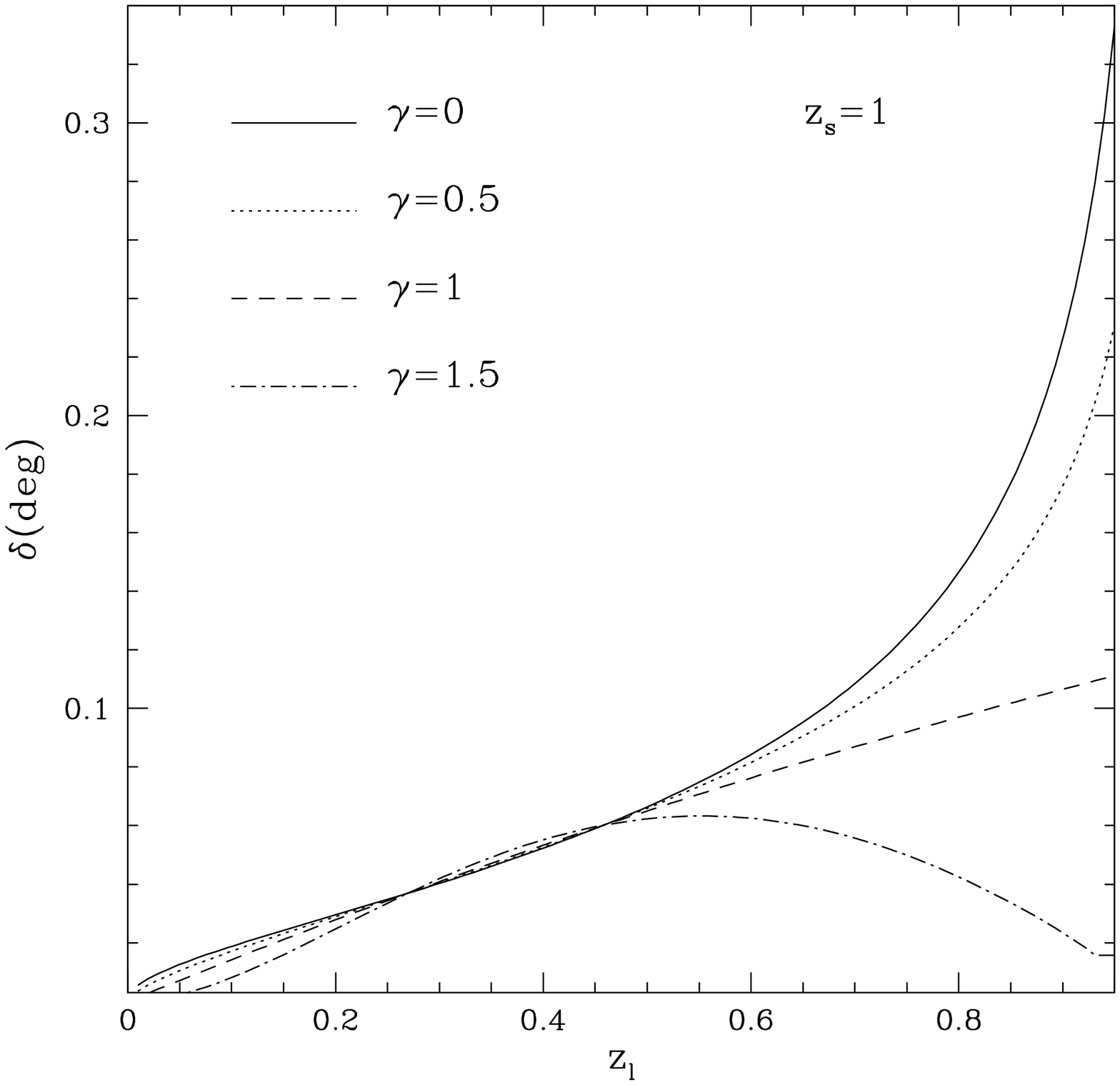}
\caption{
The angular beam separation $\delta$ for a lens with a power law
mass distribution $M=A(D_l\theta)^\gamma$, as a function of the
lens redshift.  Different curves correspond to different power law
slopes.  The point mass case ($\gamma=0$) has a mass of
$M=3\times 10^{15}M_\odot$, while in the other cases the mass
normalization $A$ is fixed so that all lenses have the same
Einstein radius when $D_{\rm l} = D_{\rm s}/2$.  The source
redshift is $z_s=6$ {\em (left panel)} and $z_s=1$
{\em (right panel)}.  We fix the source position
$\beta=10^{-4}$ deg, but note that for $\gamma\le 1$ the
dependence of $\delta$ on $\beta$ is negligible.
}
\label{fig:PL1}
\end{figure*}

\reffig{PL1} shows the results for other power law profiles,
including one that is extended but steeper than isothermal
($\gamma=0.5$), the isothermal profile ($\gamma=1$), and one
shallower profile ($\gamma=1.5$).  For all cases we fix 
$\beta=10^{-4}$ deg, which is small enough that 
we are always looking at situations with multiple images.
As we shall see, for shallow profiles the angular beam separation
is not very sensitive to the choice of $\beta$.

Qualitatively, steep profiles ($\gamma<1$) behave in a similar way
as the point mass case, with $\delta$ increasing monotonically with
$z_{\rm l}$ and diverging as the lens nears the source.  In these
cases we find that $\delta$ increases with the source position $\beta$
(the case $\gamma=0$ is shown in \reffig{PM1}; the case
$\gamma=0.5$ is not shown).

The isothermal profile ($\gamma=1$) represents a transition case,
with $\delta$ increasing monotonically with the lens redshift and
reaching a finite value as $z_{\rm l} \to z_{\rm s}$.  This is
another case we can understand analytically.  For
$\beta < \theta_E$ there are two images at positions
$\theta_{1,2}=\theta_E\pm\beta$. Combining these with the deflection
angles $\hat{\alpha}_{1,2}=(D_{\rm s}/D_{\rm ls})\,\theta_E$ yields
the angular beam separation
\beq
\delta_{\rm SIS}=2\theta_E\left(\frac{D_{\rm s}}{D_{\rm ls}} -1 \right)\;. 
\label{eq:delSIS}
\eeq 
An SIS has $\theta_E\propto D_{\rm ls}$, so when
$z_{\rm l}\rightarrow z_{\rm s}$ we see that $\delta$ approaches
a constant.  Another general point is that $\delta$ is independent
of $\beta$, while depending linearly on the mass of the 
lens ($\delta_{\rm SIS} \propto \theta_E \propto M$, where we
are considering the mass within some fixed physical radius; cf.\ 
\S\ref{sec:PL1}).

Profiles shallower than isothermal ($\gamma>1$) show a qualitatively
different behaviour: as $z_{\rm l}$ increases, the angular beam
separation $\delta$ initially rises but then reaches a peak before
turning over and returning to zero as $z_{\rm l} \to z_{\rm s}$.
While there is no simple, general expression for $\delta$ as a
function of both the power law slope $\gamma$ and source position
$\beta$, we can find an enlightening
analytic result for the limit $\beta \ll \theta_E$.  In this
case the images are near the Einstein radius,
$\theta_{1,2}\approx\theta_E$, so the two deflections angles
are $\hat{\alpha}_{1,2}\approx D_{\rm s}/D_{\rm ls}\theta_E$,
and we have $\delta\approx 2\theta_E(D_{\rm s}/D_{\rm ls}-1)$.
Then, for $D_{\rm ls} \to 0$ we have
$\theta_E\propto D_{\rm ls}^{1/(2-\gamma)}$ and hence
\beq
\delta \propto D_{\rm ls}^{1/(2-\gamma)}
  \left(\frac{D_{\rm s}}{D_{\rm ls}} -1 \right)
  \to D_{\rm ls}^{(\gamma-1)/(\gamma-2)} .
\eeq
This result clarifies the distinction between steep and shallow
profiles: as the lens nears the source, $\delta$ diverges for all
$\gamma<1$, approaches a constant for $\gamma=1$, and vanishes for
all $\gamma>1$.  Finally note that in this limit when $\beta$ is
small we have the following scaling with lens mass:
$\delta \propto \theta_E \propto M^{1/(2-\gamma)}$.

While we have explicitly shown results for a single value of the
angular source position $\beta$, we find that the general behaviour
of $\delta$ with lens redshift and mass holds for other values as
well.  We have demonstrated this analytically for the cases of
$\gamma=0$ and $\gamma=1$.  More generally, for steep profiles
($\gamma<1$), which have two images for all $\beta$, we find
that the rise of $\delta$ becomes increasingly steep as $\beta$
becomes larger than $\theta_E$.  For shallow profiles ($\gamma>1$),
by contrast, there are two images only when $\beta$ is sufficiently
small that the source lies inside the lens caustic, and it turns
out that the dependence of $\delta$ on $\beta$ is quite weak.

\section{Navarro-Frank-White mass distribution}
\label{sec:NFW}

\subsection{The lens model}
\label{sec:NFW1}

The NFW profile (Navarro, Frank \& White 1996),
\beq
\rho(r) = \frac{\rho_s}{(r/r_s)(1+r/r_s)^2}\;, 
\eeq
is fully specified by two parameters, written here as a scale
radius $r_s$ and characteristic density $\rho_s$.  It is
customary to trade these two parameters for the (virial) mass
$M_{\rm vir}$ and concentration $c$.  The transformation goes by
way of the virial radius $r_{\rm vir}$, which is related to the
virial mass by
$M_{\rm vir} = (4\pi/3)\,\Delta_{\rm vir}\,{\bar\rho}\,r_{\rm vir}^3$,
where $\Delta_{\rm vir}$ is the virial overdensity, and the
mean matter density is ${\bar\rho}(z) = 3H_0^2\Omega_m(1+z)^3/8\pi G$.
For the virial overdensity we use the fitting formula provided
by Bullock et al. (2001):
$\Delta_{\rm vir}\approx (18\pi^2+82 x - 39 x^2)/\Omega(z)$,
where $x\equiv\Omega(z)-1$, and $\Omega(z)$ is the ratio of the
mean matter density to the critical density at redshift $z$.
Using these ingredients, we can finally specify the original
NFW parameters: $r_s = r_{\rm vir}/c$, and
$\rho_s = \delta_c\,{\bar\rho}$, and
\beq
\delta_c(z) = \frac{\Delta_{\rm vir}(z)}{3}\frac{c^3}{[\ln(1+c)-c/(1+c)]}\;.
\eeq 

\begin{figure*}
\includegraphics[width=0.45\textwidth]{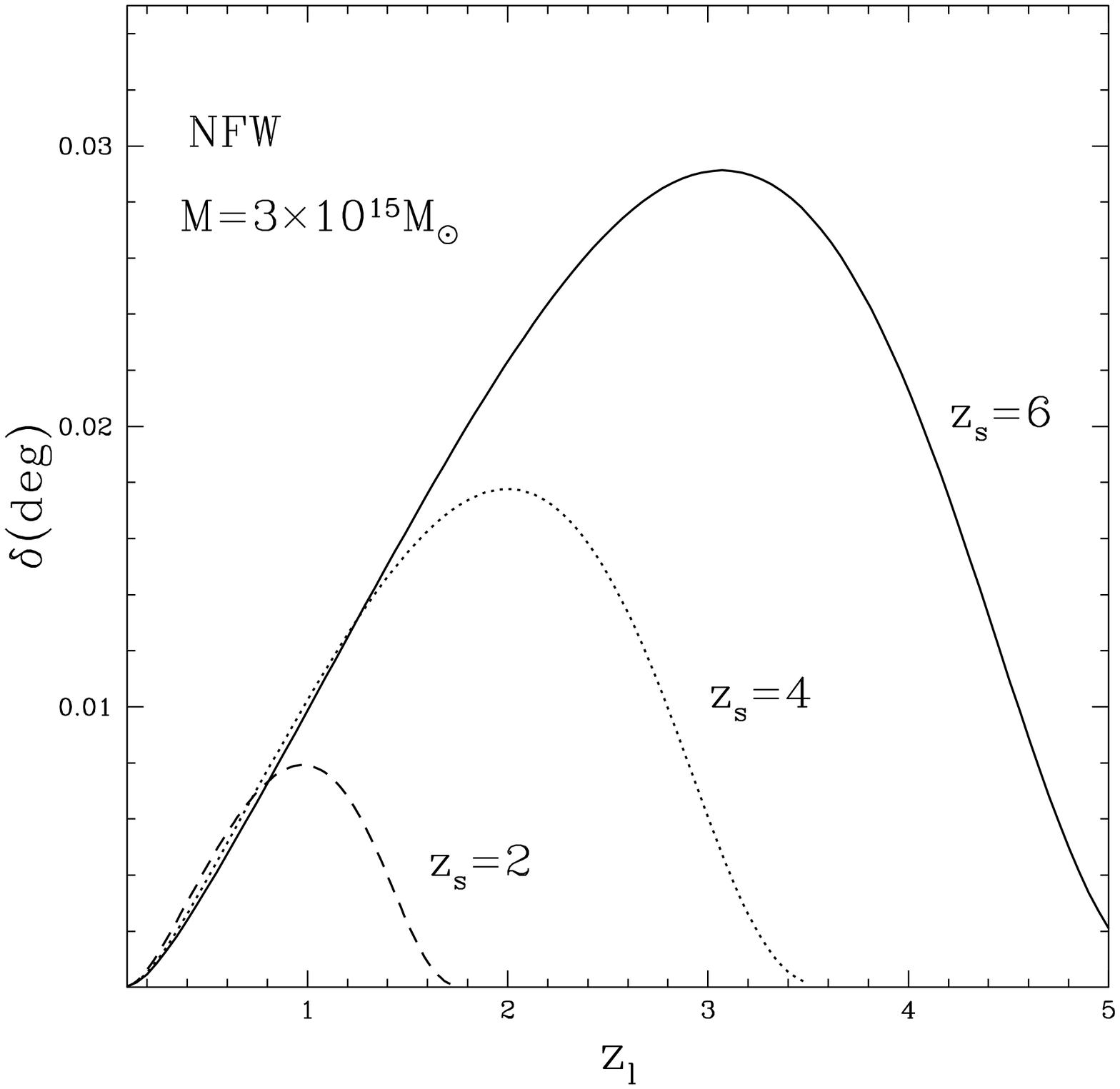}
\includegraphics[width=0.45\textwidth]{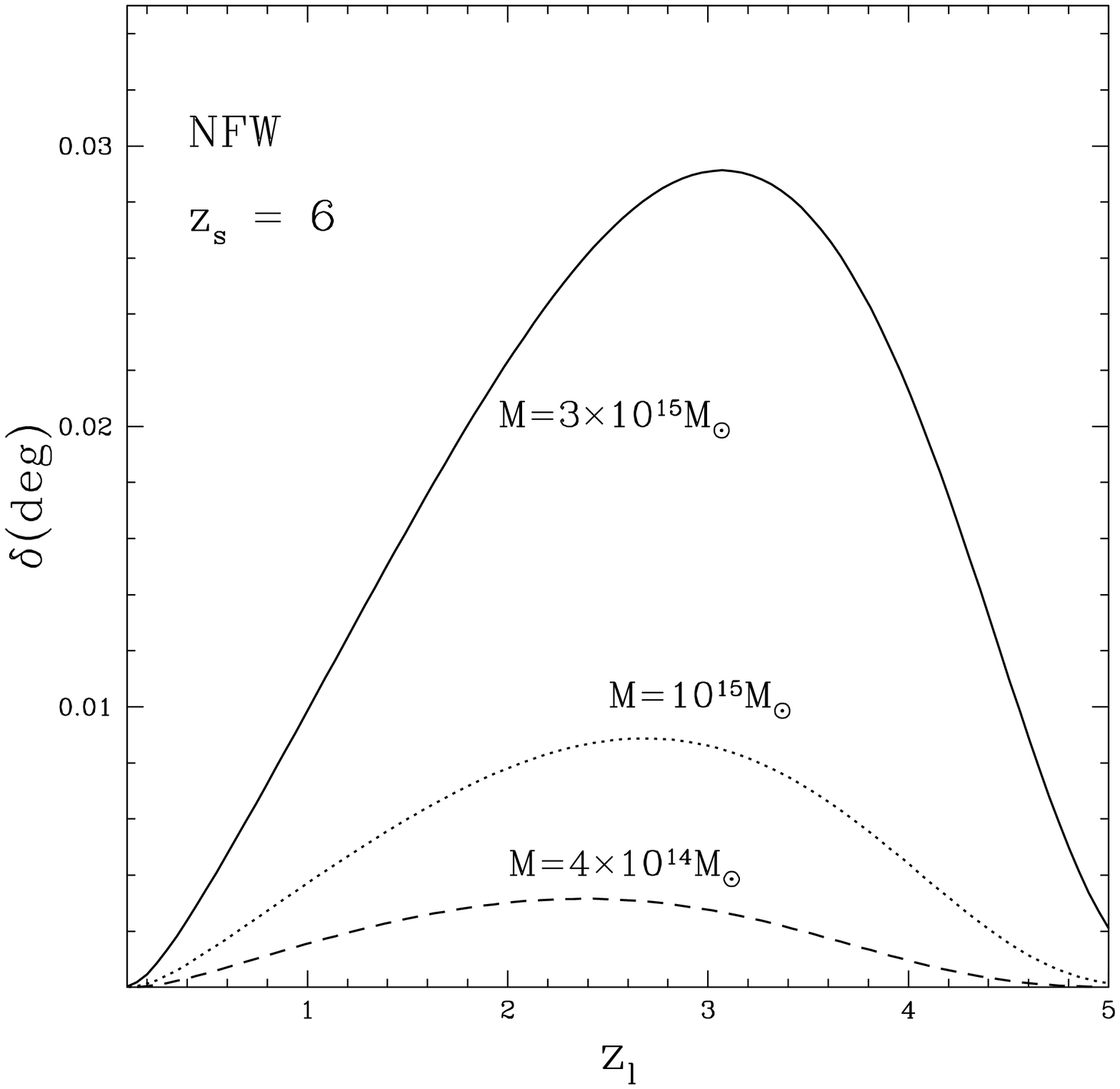}
\caption{
The angular beam separation $\delta$ as a function of lens redshift
for an NFW lens with three different source redshifts {\em (left panel)}
and three different lens masses {\em (right panel)}.  The source
position is taken to be $\beta=0.05\beta_{\rm caus}$, but $\delta$
is not very sensitive to this choice.
}
\label{fig:NFWz}
\end{figure*}

The two NFW parameters are not entirely independent: numerical
simulations reveal a relation between the mass and concentration
of NFW halos.  Bullock et al. (2001) describe this relation as a
log-normal distribution for the concentration whose median value
depends on the halo mass and redshift,
\beq
c_{\rm med} (M,z)=\frac{\bar{c}_s}{1+z}\left(\frac{M}{M_*}\right)^{\alpha},
\eeq
where $M_*$ is the mass of a typical halo collapsing today.  For
a cosmology with $\sigma_8=0.9$ and $w=-1$,
$M_*=1.13\times 10^{13}M_{\odot}$ (Kuhlen, Keeton \& Madau 2004).
The halos in the $\Lambda$CDM simulations by Bullock et al. were
best described by the parameter values $\bar{c}_s=9.0$ and
$\alpha=-0.13$.  Although the simulations show a scatter, for
simplicity we just use the median concentration at each mass and
redshift.

The lensing characteristics of an NFW halo are given by
Bartelmann (1996).  The projected surface mass density, scaled
by the critical density for lensing, is
\beq
\kappa(x)=2\kappa_s \;\frac{f(x)}{x^2-1}\;,
\label{eq:kr}
\eeq
where 
$x\equiv r/r_s$,
\beq
f(x)=\left\{
\begin{array}{ll}
1-\frac{2}{\sqrt{x^2-1}}\arctan\sqrt{\frac{x-1}{x+1}}        &\hbox{if $x>1$}\\
1-\frac{2}{\sqrt{1-x^2}}\mbox{arctanh}\sqrt{\frac{1-x}{x+1}} &\hbox{if $x<1$}\\
0 & \hbox{if $x=1$}
\end{array}\right.
\label{eq:fx}
\eeq
and $\kappa_s\equiv {\rho_s r_s}{\Sigma_{\rm crit}^{-1}}$ is a
characteristic surface mass density in units of the critical density
for lensing,
\beq
\Sigma_{\rm crit}=\frac{c^2}{4\pi G}\frac{D_{\rm s}}{D_{\rm l} D_{\rm ls}}\;.
\eeq
The (scaled) mass inside radius $x$ is
\beq
m(x)=2\int_0^x dx' x' \kappa(x')= 4\kappa_s g(x)\;,
\label{eq:mx}
\eeq 
where $g(x)=\log(x/2) + 1 -f(x)$.  With this notation, we can write
the lens equation in scaled coordinates as
\beq
y = x - \frac{4 \kappa_s g(|x|)}{x}\ ,
\label{eq:NFWlens}
\eeq
where $y = \beta/\theta_0$ is the source angle scaled by the
angular NFW scale radius, $\theta_0 = r_s/D_{\rm l}$.  An NFW lens
has one tangential and one radial critical curve, whose scaled
radii $x_t$ and $x_r$ are given by the solutions of the equations
\beq
1-\left.\frac{m(x)}{x^2}\right|_{x=x_t}=0\;,\;\;\;\;\;1-\left.\frac{d}{dx}\frac{m(x)}{x}\right|_{x=x_r}=0\;.
\label{eq:crit}
\eeq
The tangential critical curve maps to the origin in the source plane
(i.e., it represents the Einstein ring), while the radial critical
curve maps to the caustic in the source plane, which has angular
radius $y_{\rm caus}=m(x_r)/x_r-x_r$.   If $y < y_{\rm caus}$, an
NFW lens produces three images: one of the main images is outside
$x_t$, the other is between $x_t$ and $x_r$, and the third image
(the one we ignore) is inside $x_r$.  All of these scaled variables
can be converted into angular variables by multiplying by $\theta_0$.

For an image at angular position $\theta$, the deflection angle is
\beq
  \left|{\hat{\alpha}}(\theta)\right| = \frac{D_{\rm s}}{D_{\rm ls}}\;
    \frac{{m(|\theta|/\theta_0)}}{{|\theta|/\theta_0}}\;\theta_0\;.
\eeq

\subsection{The angular beam separation}
\label{sec:NFW2}

\reffig{NFWz} shows the angular beam separation $\delta$ as a function
of lens redshift for certain values of the lens mass and source
redshift.  From a qualitative point of view, the curves resemble 
the curve for the shallow power law case in \reffig{PL1}.
This is not surprising, since the NFW profile is shallower than
isothermal inside the scale radius $r_s$, and the Einstein radius
is (much) less than the scale radius for all cases of interest.

\begin{figure*}
\includegraphics[width=0.45\textwidth]{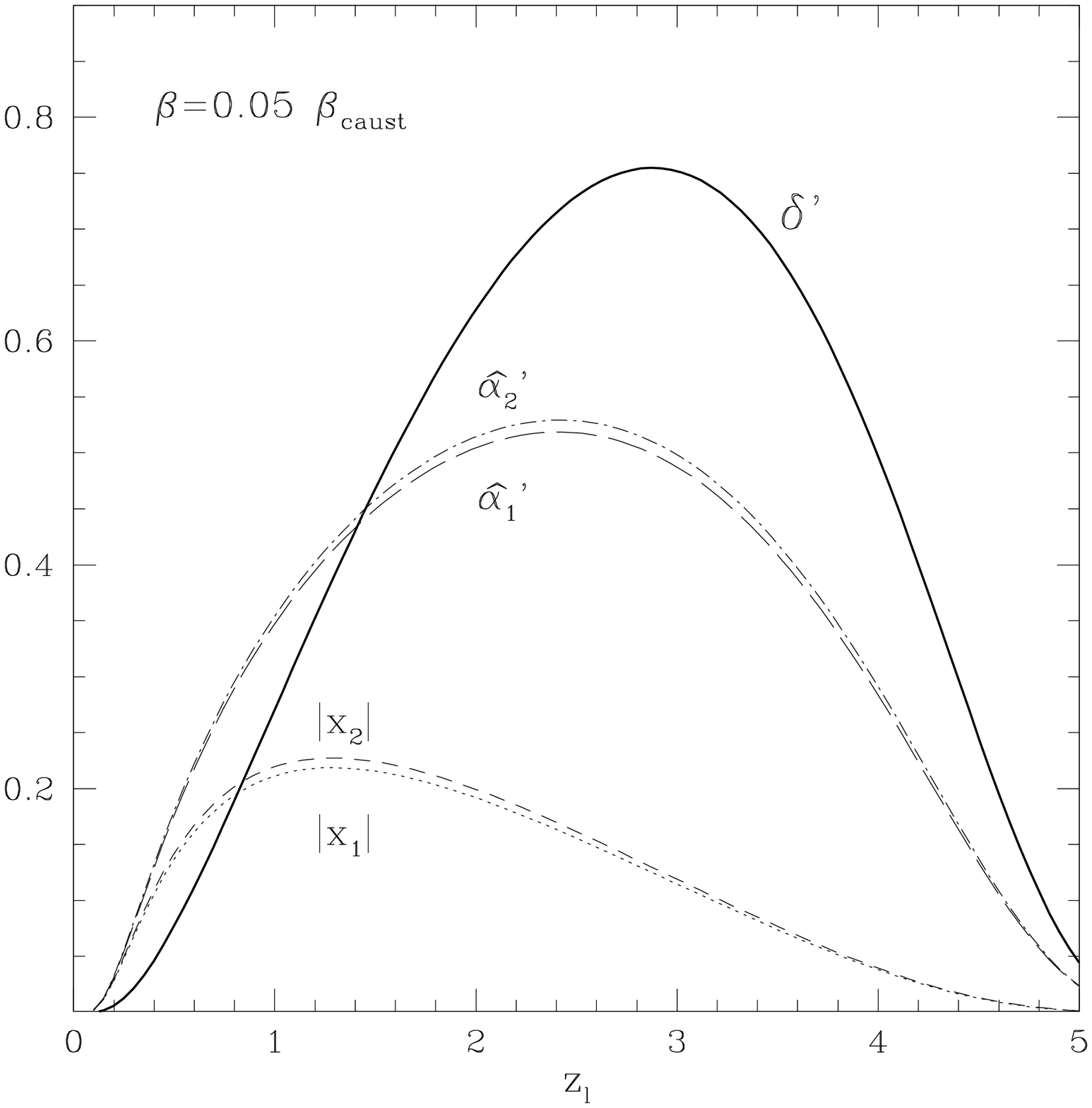}
\includegraphics[width=0.45\textwidth]{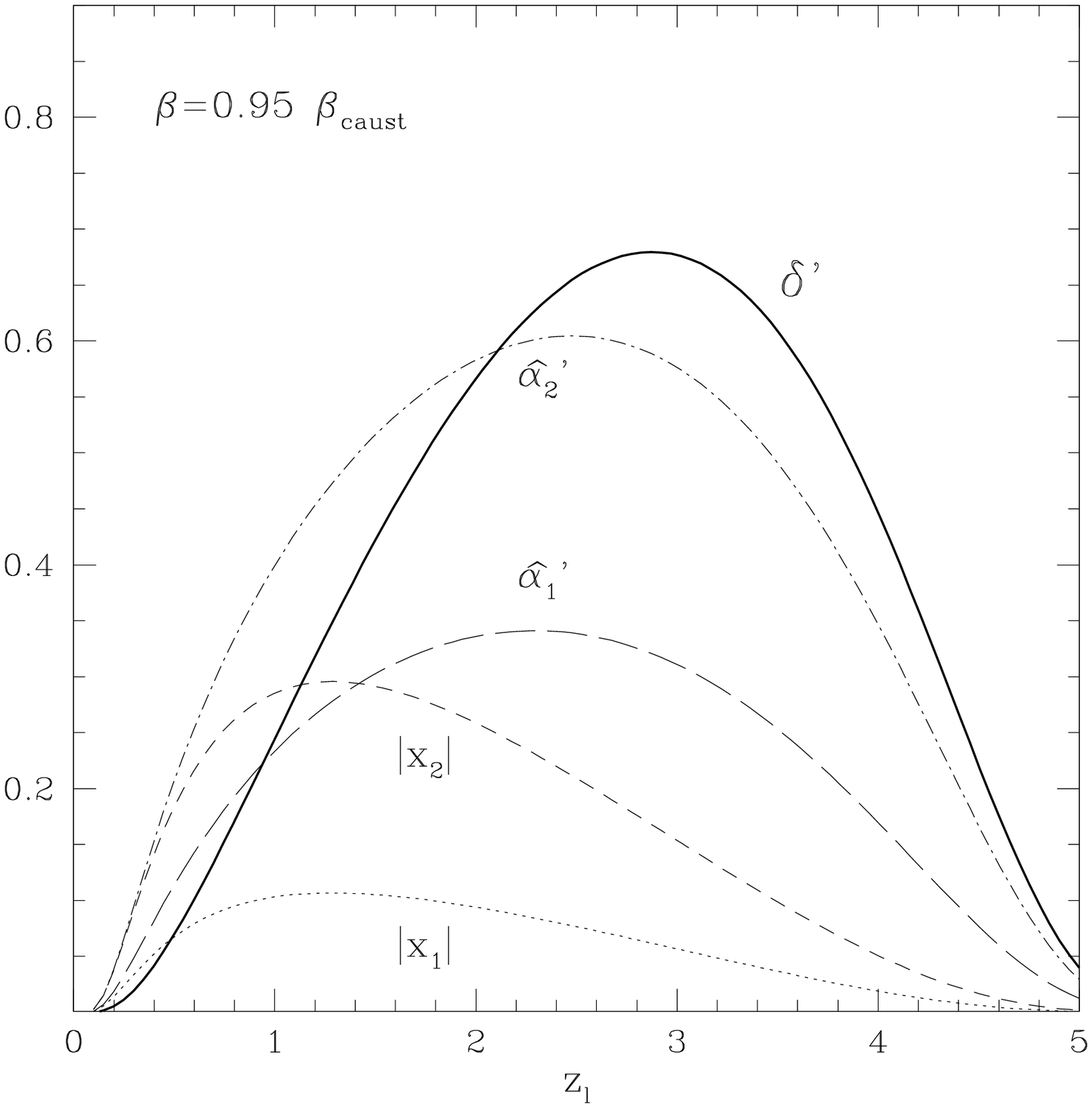}
\caption{
Image positions ($x_{1,2}$), deflections angles (${\hat\alpha}_{1,2}$)
and angular beam separation ($\delta$) for an NFW lens.  All quantities
are in units of $\theta_0$ (the 'prime' distinguishes them from the
corresponding quantities in angular units).
The lens mass is $M=3\times 10^{15} M_{\odot}$,
and the source redshift is $z_s=6$. In the left panel the source is
close to the optical axis ($\beta=0.05\beta_{\rm caus}$), while in
the right panel it is close to the caustic
($\beta=0.95\beta_{\rm caus}$). Note that the image positions and
deflection angles depend on the source position, but the angular beam
separation $\delta$ is almost independent of $\beta$.
}
\label{fig:check}
\end{figure*}

The $\delta(z_{\rm l})$ curves peak at a redshift which is about
half that of the source.  This is curious but coincidental; it
arises from the different redshift dependences of the image
positions and deflection angles.  \reffig{check} shows $\delta$
along with its four consituents: the image positions $\theta_{1,2}$
and deflection angles ${\hat\alpha}_{1,2}$.  (These are plotted
in units of $\theta_0$, but that scale factor has a weak dependence
on $z_l$.)  Neither the image position curves nor the deflection
angle curves peak at $\sim z_{\rm s}/2$; the image position curves
peak at lower redshift, in keeping with the rule of thumb that a
lens is most effective when it is about half the {\em distance}
to the source.  As far as we can tell, it is purely coincidental
that the different redshift dependences cause the peak of the
$\delta$ curve to be located at $\sim z_{\rm s}/2$.

Another interesting point from \reffig{check} is that the angular
beam separation is quite insensitive to the source position even
though its constituents do depend on $\beta$.  This is because the
image position and deflection angle depend on $\beta$ in a similar
way: as $\beta$ increases, $\theta_1$ and ${\hat\alpha}_1$ both
increase while $\theta_2$ and ${\hat\alpha}_2$ both decrease (in
amplitude).  The way these terms combine to form the angular beam
separation means the changes largely cancel and leave $\delta$
relatively insensitive to $\beta$.

\begin{figure}
\begin{center}
\includegraphics[width=0.45\textwidth]{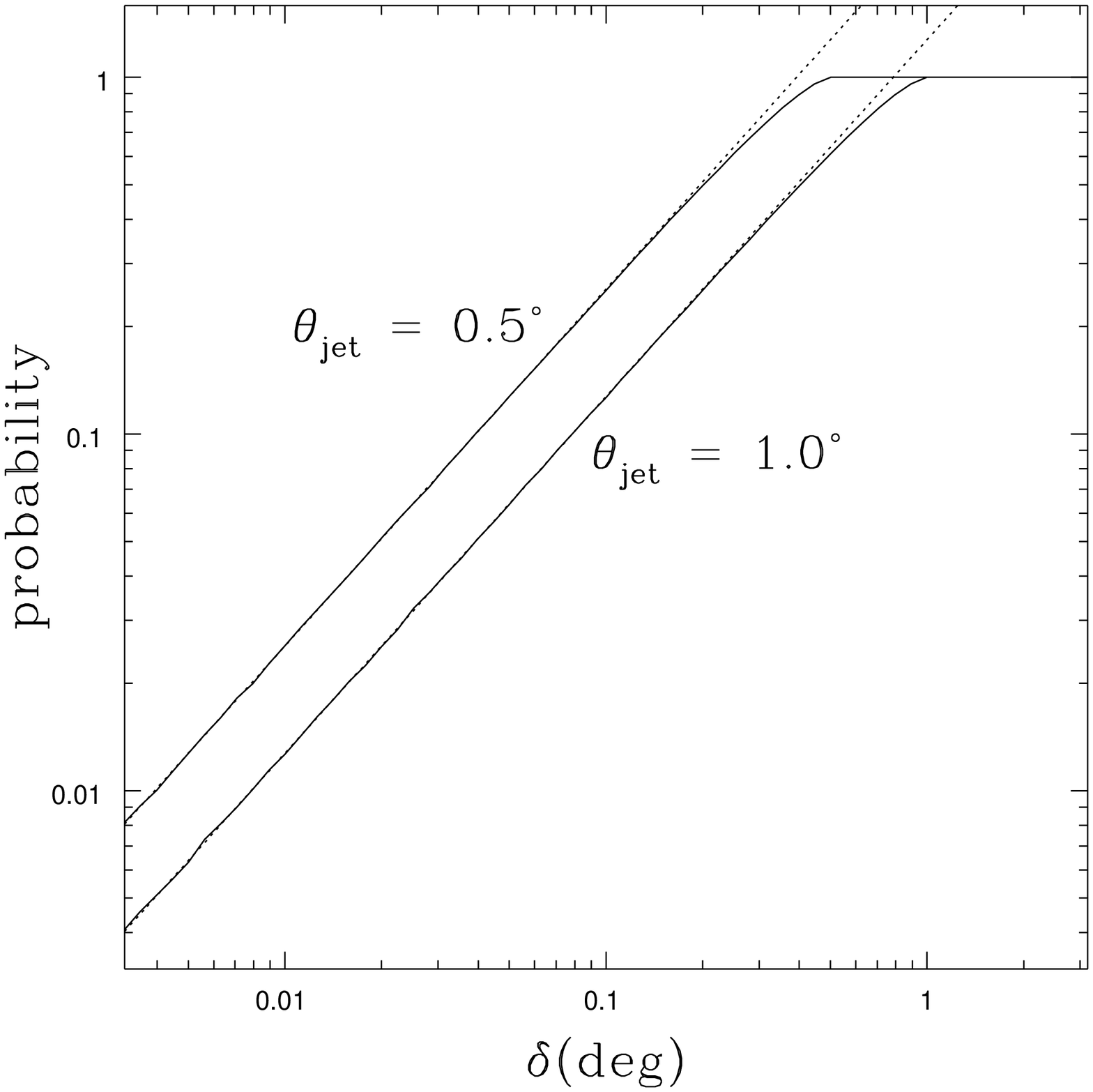}
\end{center}
\caption{
For a beamed source with opening angle $\theta_{\rm jet}$, the solid
curves show the probability that one of the lensed images is missed,
given that at least one image is seen, as a function of the angular
beam separation $\delta$.  The probability saturates at unity for
$\delta \ge \theta_{\rm jet}$ (see text).  The dotted lines show the
approximation $P \approx (4\delta)/(\pi\theta_{\rm jet})$.
}
\label{fig:Pjet}
\end{figure}

To illustrate one way of interpreting the $\delta$ values, let us
consider lensing of a beamed source.  Depending on the angular beam
separation $\delta$ and the jet opening angle\footnote{We take
$\theta_{\rm jet}$ to be the full opening angle, i.e., from one
side of the jet to the other.} $\theta_{\rm jet}$, there may be
configurations in which we see only a single image even though
conventional lens theory (which assumes an isotropic source) would
predict two.  For example, in the limit that the angular beam
separation is larger than the jet opening angle
($\delta > \theta_{\rm jet}$), there is no way to arrange the jet
such that we could see both images: we could see one image or the
other, or nothing at all (if the jet does not point along either
light ray), but never both images.  In this case the probability
that we would ``miss'' one of the images predicted by conventional
lens theory is unity.

For $\delta < \theta_{\rm jet}$ there is some finite probability of
missing one of the images.  Given $\delta$ and $\theta_{\rm jet}$,
we use simple numerical simulations to consider all possible jet
orientations and compute the conditional probability that one image
is missed, given that at least one image is seen.  (The conditional
part of the probability just means we do not consider cases where
the jet is pointing ``away'' from us so that we cannot see anything.)
The results are shown in \reffig{Pjet}.  In the limit
$\delta \ll \theta_{\rm jet} \ll 1$ we can derive a useful analytic
approximation $P \approx (4\delta)/(\pi\theta_{\rm jet})$, which is
also shown in the figure.  Combining the probability results with
the $\delta$ values from \reffig{NFWz}, we deduce that there is some
finite probability that a beamed source could be ``lensed'' by a
massive NFW cluster in such a way that we miss one of the images,
and that this probability could be as high as
$P \sim 0.02$--$0.07\,(\theta_{\rm jet}/{0.5}^\circ)^{-1}$
depending on the redshift of the source.  This probability is
largest when $z_{\rm l} \sim z_{\rm s}/2$.  (By contrast, the
corresponding probability for a steep power law mass profile
would become ever larger as $z_{\rm l} \to z_{\rm s}$.)

\section{Cosmological population of NFW lenses}
\label{sec:MC}

The effects of source anisotropy depend on the angular beam separation,
which in turn depend on the redshift, mass, and profile of the lens.
For a lens system in which the lensing galaxy or cluster is known,
it would be natural to study source anisotropy using the specific lens
configuration.  But for a system in which the lensing object has not
yet been identified, it would be important to understand the statistical
distribution of angular beam separations.  We now examine that
distribution using Monte Carlo simulations of a population of halos
between the observer and source.  Since we have seen that the angular
beam separation is largest for massive lenses (which is not surprising),
we focus on cluster lenses here.  We treat the clusters using the NFW
profile, which of course is oversimplified but serves to give useful
first estimates.

We model the cluster population using the mass function from
Warren et al. (2006),
\begin{eqnarray}
n(M,z) &=& 0.7234\left(\sigma_M^{-1.625}+0.2538\right)
  e^{-1.1982/\sigma_M^2} \nonumber\\
&& \times \frac{{\hat\rho}(z)}{M^2}\ 
  \frac{{\rm d}\log \sigma_M^{-1}}{{\rm d}\log M}\;,
\label{eq:warren}
\end{eqnarray}
where ${\hat\rho}(z)$ is the mean comoving matter density at redshift
$z$, and $\sigma_M$ is the linear density fluctuation on mass scale
$M$.  In terms of the power spectrum,
$\sigma^2_M=(1/2\pi^2)\int dk\,k^2\,P(k)\,W(kR)^2$,
where $P(k)$ is the matter power spectrum, $W(kR)$ is the window
function corresponding the smoothing of the density field
(e.g., Peebles 1993), and $R$ is the comoving scale corresponding
to a mass $M=(4\pi/3){\bar\rho} R^3$.

Let us now consider a source at redshift $z_s$.  The probability
that its light is (multiply) lensed by a mass on its way to the
observer, is given by the fraction of the sky that is covered by
lens caustics:\footnote{If we were considering a particular
flux-limited survey, we would need to consider not only the fraction
of the sky covered by caustics (the lensing optical depth) but also
the fact that lensing configurations with higher magnifications are
easier to detect (magnification bias).  We find, though, that
magnification bias does not significantly affect our conclusions
about the distribution of $\delta$ values.}
\beq
  P_{\rm lens} = \frac{1}{4\pi}\int \int f(M,z) \, dM \,dz\;,
  \label{eq:plens}
\eeq
where
\beq
f(M,z)= \pi \beta_{caus}^2\; n(M,z)\; \frac{dV}{dz}(z)\;,
\label{eq:fun}
\eeq
and $dV/dz(z)$ is the comoving volume at redshift $z$. 
We can interpret $f(M,z)$ (once it is normalized) as the joint
probability distribution for the halo mass and redshift, and draw
from this distribution as follows.  First, we obtain the redshift
distribution by marginalizing over mass:
$P'_{\rm lens}(z)=A_1 \int_{M_{\rm min}}^{M_{\rm max}} f(M,z)\,dM$,
where $A_1$ is a normalization factor.  We consider the mass range
from $M_{\rm min}=10^{14} M_\odot$ to
$M_{\rm max}=3\times 10^{15} M_\odot$.  As we shall see, reducing
the minimum mass would principally affect the distribution of
angular beam separations at the small-$\delta$ end, which is not
so interesting from the standpoint of expected anisotropy scales.
Our results are not very sensitive to the upper mass cut because
higher mass halos are exponentially rare.  We draw a random lens
redshift $z_{\rm l}$ from $P'_{\rm lens}(z)$, and then construct
the conditional probability distribution for the mass:
$P''_{\rm lens}(M|z_{\rm l})=A_2\,f(M,z_{\rm l})$, where $A_2$
is again a normalization factor.  Note that the factor of
$\pi \beta_{\rm caus}^2$ in eq.~(\ref{eq:fun}) ensures that each
halo is weighted by its lensing cross section.

Once we have drawn a lens mass and redshift, we need to draw a
random source position.  Since we consider only multiply-imaged
sources, we can restrict attention to
$0 \le \beta \le \beta_{\rm caus}$.  We assume a uniform distribution
of source positions (in 2-d), which is equivalent to
$P(\beta)\,d\beta = (2\beta/\beta_{\rm caus}^2)\,d\beta$.  We note
that lensing magnification bias can cause the distribution of source
positions to be non-uniform.  However, since the angular beam
separation for NFW lenses is not very sensitive to $\beta$
(even for $\beta$ on or near a caustic; cf.\ \S\ref{sec:NFW2}),
our results are not very sensitive to the distribution of source
positions.

\begin{figure*}
\includegraphics[width=0.45\textwidth]{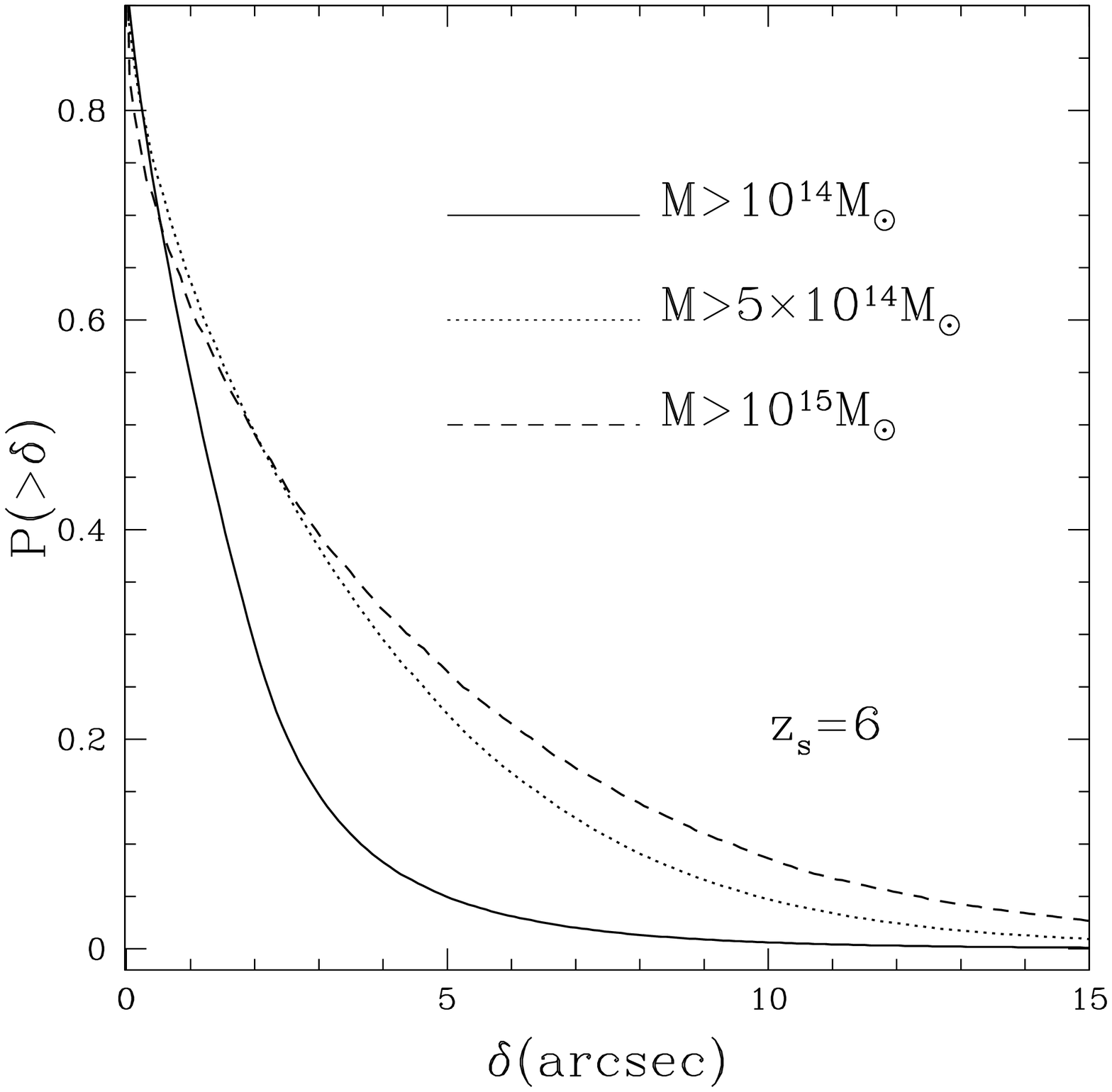}
\includegraphics[width=0.45\textwidth]{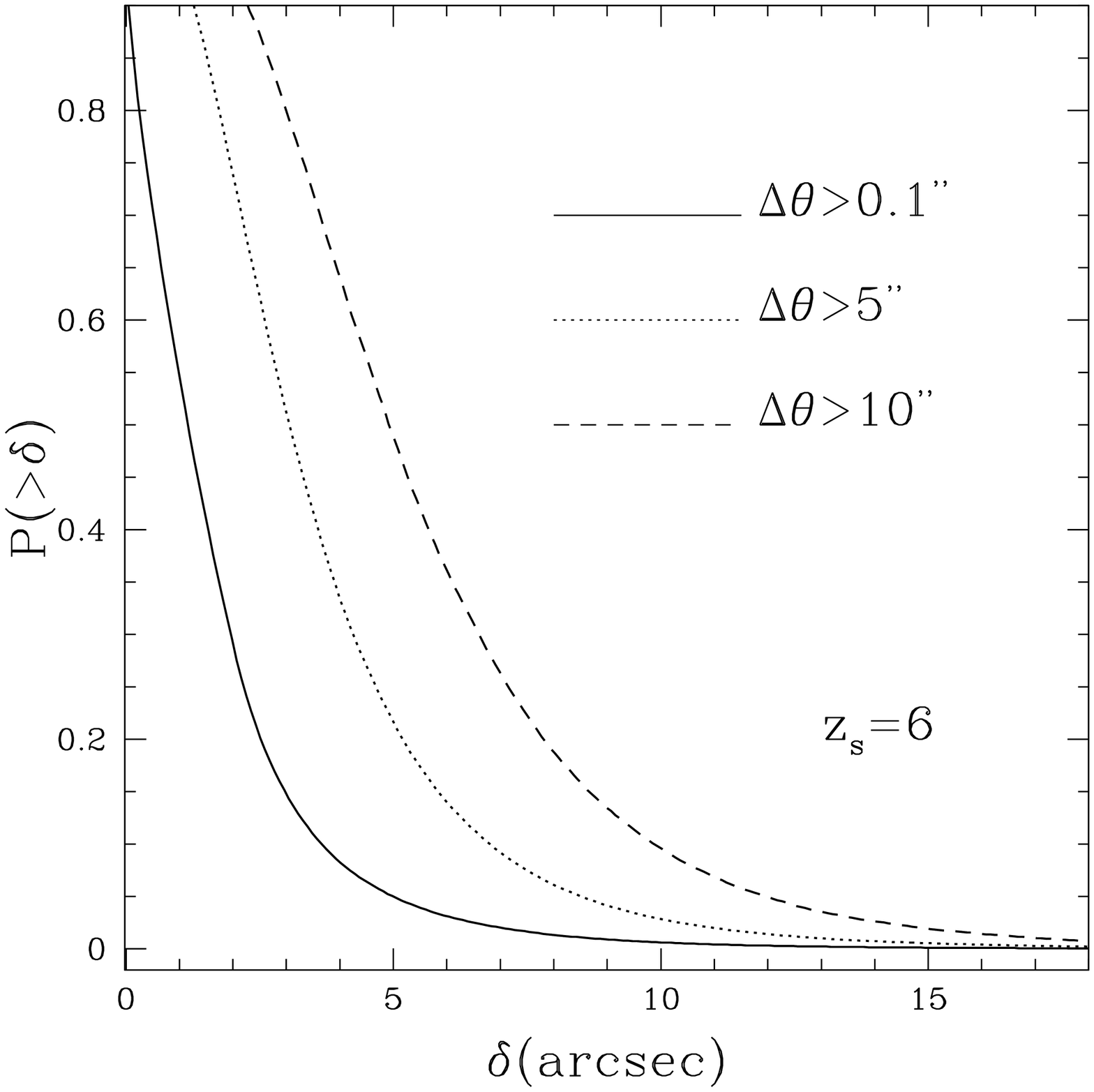}
\caption{
Cumulative probability distribution for a lensed source at redshift
$z_s = 6$ to have an angular beam separation greater than $\delta$.
The left panel shows results for different mass cuts, while the right
panel shows results for different image separation cuts.
}
\label{fig:delta}
\end{figure*}

For each set of $(z_{\rm l}, M, \beta)$ values we compute the angular
beam separation $\delta$, and then repeat the process many times to
obtain the $\delta$ distribution.  \reffig{delta} shows the results.
In analyzing the results, we examine one case in which we consider all
halos above a fixed mass theshold (left panel of \reffig{delta}),
and a second case in which we consider all halos that produce a lens
image separation larger than some value (right panel).

The range of $\delta$ values typical for the statistical distribution
is somewhat smaller than the values seen in \reffig{NFWz}, which is
not surprising because the cluster population tends to be dominated
by clusters at lower masses and redshifts than were used for the
previous example, and both effects tend to reduce $\delta$.  A useful
corollary is that, since $\delta(z_{\rm l})$ is largely insensitive
to $z_{\rm s}$ (see the left panel of \reffig{NFWz}), the $\delta$
distributions shown in \reffig{delta} are not very dependent upon
the assumed source redshift.

The typical value of $\delta$ increases with the mass or image
separation threshold, which is also not surprising. If we consider
wide-separation lenses with $\Delta\theta > 10\arcsec$, the median
angular beam separation is $\delta \approx 5\arcsec$, and there is
about a 10\% chance that the angular beam separation would be larger
than about $10\arcsec$.  As discussed below, this range of $\delta$
values corresponds to possible scales of anisotropy in real
astrophysical sources, which suggests that source anisotropy may have
non-negligible if not dominant effects in strong lensing.

\section{Summary and Discussion}
\label{sec:summary}

Motivated by the observation that anisotropic emission is ubiquitous
in astrophysical sources, we have addressed the general problem of
gravitational lensing of an anisotropic source.  For simplicity, in
this pilot study we have focused on spherical mass distributions. 

The slope of the mass profile plays a crucial role in determining
the angular beam separation $\delta$ between the two main lensed
images, and hence the probability that the appearance of the images
will be affected by source anisotropy (that we might miss one of the
images of a beamed source, for example).  If the mass profile is
steeper than isothermal, $\delta$ increases with the redshift of
the lens (for a fixed source redshift).  If the mass profile is
shallower than isothermal, $\delta$ peaks when the lens is somewhere
in between the observer and the source.  In the case of an NFW lens,
the peak occurs for $z_{\rm l}\sim z_{\rm s}/2$.   In general,
the angular beam separation $\delta$ is not necessarily similar to
the image separation, $\Delta\theta$.

By modeling a cosmological distribution of clusters with NFW
profiles, we have estimated the distribution of angular beam
separations---and, by implication, the range of source anisotropy
scales that are most likely to affect lensing.  Typical values of
$\delta$ for cluster lenses lie in the range of a few to tens of
arcseconds.  We remark that while our statistical analysis should
be instructive, there are some complications we have omitted in
this pilot study.  If baryons steepen the central density profile
(compared with the pure NFW profiles we have considered; see
Puchwein et al. 2005; Rozo et al. 2008), that would tend to
increase the $\delta$ values.  In general we would expect
asphericity to change $\delta$ only by a factor of order unity,
although that would introduce some new phenomenology in terms of
how source anisotropy would affect lenses with more than two bright
images corresponding to more than two light rays emerging from the
source.

Our results have implications for a range of astrophysical
sources. Gamma-Ray Bursts are known to be extremely relativistic
sources, with Lorentz factors $\Gamma\sim 100$--$300$. This implies
that each point on the emitting surface is only visible to observers 
within a sub-degree scale.  If the GRB emitting region is confined
within a region $\la 1^\circ$ (as suggested by recent numerical
simulations; Komissarov et al. 2009), GRBs would be excellent
candidates to display effects of anisotropic lensing.  Since GRBs
are short-lived sources, GRB lensing would not generally result in
contemporaneous  multiple imaging.  Rather, if a GRB goes off behind
a massive cluster, one would naively expect to see a nearly
identical (modulo some magnification factor) GRB within the same
instrumental error circle some months or years later.  We have
found, however, that there could be a small but finite probability
that one of the images could be missed because of the angular beam
separation between the light rays that form the lensed images.
As a specific example, if we consider a GRB at $z_{\rm s} = 6$
lensed by a massive cluster at $z_{\rm l} = 1$ (i.e., the scenario
depicted in \reffig{NFWz}), and its jet opening angle is
$\theta_{\rm jet} = 1^\circ$, there is a 1\% probability that the
second GRB image will not appear.  That probability doubles if
$\theta_{\rm jet} = 0.5^\circ$, and it increases to about 6\%
for a lens redshift $z_{\rm l} = 3$.  The possibility of missing
images ought to be incorporated into statistical forecasts of
GRB lensing (which have heretofore assumed isotropic source
emission; cf.  Porciani \& Madau 2001).

Other sources that have been suggested to be highly relativistic are
blazars. Giannios et al. (2009) suggested that the fast TeV
variability observed in two sources can be explained as the result of
compact emitting regions moving towards the observer with Lorentz
factors of $\sim 100$, and embedded within a jet moving at lower
speed.  In this scenario, the emission from each blob is beamed within
a sub-degree scale. Depending on the angle at which the blob is moving
with respect to the line of sight to the observer, there is some
probability that a lensed blazar might be missing one of the
images. If that does happen to be the case in lensing observations
(e.g., when a blazar is observed behind a large cluster and the
number of images appears anomalous) it would provide support for
this physical picture of blazars.

Anisotropy in the net flux leaving the source can also result from
inhomogeneous absorption within the source.  This is indeed the
case for AGNs, in which dense clouds in the broad abosorption line
region create a highly anisotropic absorbtion pattern.  The precise
location and size of these clouds is still a controversial issue.
Estimates suggest that their size is not larger than about
$10^{14}$ cm (e.g., Baldwin et al. 1995; Elvis 2000), while their
location has been placed in a range between 0.01 and 1000 pc (e.g.,
de Kool et al. 2001; Everett et al. 2002). These scales fall in a
quite interesting range for our problem.  A cloud of size $10^{14}$ cm
at a distance of 1 pc would subtend an angle of about $7\arcsec$,
which is similar to the median angular beam separation for NFW
lenses with image separations $\Delta\theta > 10\arcsec$.  If the
cloud were significantly closer to the central engine than 1 pc, it
would most likely cover both of the light rays that correspond to
lensed images, so the importance of source anisotropy would depend
on whether there is significant internal structure within BAL clouds
on scales smaller than $10^{14}$ cm.  Conversely, if BAL clouds are
significantly farther than 1 pc, the importance of source anisotropy
would depend on the covering fraction of BAL clouds.

If an AGN is strongly lensed and there is significant differential
absorption within the source, that would effectively cause different
lensed images to have different source fluxes, which would in turn
break the connection between observed flux ratios and lensing
magnification ratios.  If this complication is not recognized, it
could lead to errors in lens models and their interpretation.  By
contrast, if the differential source absorption is recognized, the
ability to simultaneously probe multiple lines of sight into the
source with strong lensing would provide a new way to probe the
structure of the absorbing medium in AGN, which is still very
uncertain.  This possibility is related to the suggestion by
Chelouche (2003) and Green (2006) that lensed quasars can be
used to study small-scale structure in quasar outflows.  One good
would to identify differential absorption would be to compare flux
measurements at both X-ray and optical wavelengths (Green 2006).
Column densities inferred for the absorbers in the broad line
regions are $N_{\rm H}\ga 10^{22}$ cm$^{-2}$, much larger than
derived from the UV (Green et al. 2001; Gallagher et al. 2002).
Therefore, one expects that a light ray passing through an absorber
would have a smaller X-ray/optical flux ratio than a non-absorbed
ray.

In summary, anisotropy in sources that are gravitationally lensed
could influence the appearance of the lensed images---including
whether we even see all the images.  The effect will be most
significant for wide-separation lenses produced by cluster-mass
objects.  Source anisotropy will probably not dramatically alter
the statistics of GRBs, blazars, and AGNs lensed by clusters,
but its effect may be non-negligible and certainly ought to be
considered.  If effects of source anisotropy can be recognized,
they would provide a unique opportunity to learn more about the
small-scale structure of the emitting region of the source.

\section*{Acknowledgements}

We thank Nahum Arav for discussions on absorbers in AGNs, Elena
Pierpaoli for discussions on clusters, and Kelly Wieand for
discussions about the jet probability calculation.  CRK acknowledges
support from NSF through grant AST-0747311.


\begin{thebibliography}{}

\bibitem{} Baldwin, J., Ferland, G., Korista, K., Verner, D. 1995, ApJ, 455, L119
\bibitem{} Bartelmann, M. 1996, A\&A, 313, 697
\bibitem{} Bullock,  J. S., Kolatt, T. S., Sigad, Y., Somerville, R. S.,
Kravtsov, A. V., Klypin, A. A., Primack, J. R., Dekel, A. 2001, MNRAS, 321, 559
\bibitem{} Burke, W. L., 1981, ApJ, 244, L1
\bibitem{} Chelouche, D. 2003, ApJ, 596, L43
\bibitem{} Chiba, M. 2002, ApJ, 565, 17
\bibitem{} Chiba, M., Minezaki, T., Kashikawa, N., Kataza, H., \& Inoue, K. T. 2005, ApJ, 627, 53
\bibitem{} Dalal, N., \& Kochanek, C. S. 2002, ApJ, 572, 25
\bibitem{} de Kool, M., Arav, N., Becker, R. H., Gregg, M. D., White, R. L., Laurent-Muehleisen, S. A., Price, T., Korista, K. T. 2001, ApJ, 548, 609
\bibitem{} Eigenbrod, A., Courbin, F., Vuissoz, C., Meylan, G., Saha, P., \& Dye, S. 2005, A\&A, 436, 25
\bibitem{} Elvis, M. 2000, ApJ, 545, 63
\bibitem{} Everett, J. Konigl, A., \& Arav, N. 2002, ApJ, 569, 671
\bibitem{} Fohlmeister, J., et al. 2007, ApJ, 662, 62
\bibitem{} Fohlmeister, J., Kochanek, C. S., Falco, E. E., Morgan, C. W., \& Wambsganss, J. 2008, ApJ, 676, 761
\bibitem{} Gallagher, S. C., Brandt, W. N., Wills, B. J., Charlton, J. C., Chartas, G.,
Laor, A. 2004, ApJ, 603, 425
\bibitem{} Green, P. J. et al. 2001, ApJ, 558, 109
\bibitem{} Green, P. J. 2006, ApJ, 644, 733
\bibitem{} Grossman, S. A., \& Nowak, M. A. 1994, ApJ, 435, 548
\bibitem{} Holz, D. E., Miller, M. C., \& Quashnock, J. M. 1999, ApJ, 510, 54
\bibitem{} Keeton, C. R., Gaudi, B. S., \& Petters, A. O. 2003, ApJ, 598, 138
\bibitem{} Keeton, C. R., Gaudi, B. S., \& Petters, A. O. 2005, ApJ, 635, 35
\bibitem{} Keeton, C. R., \& Madau, P. 2001, ApJ, 549, L25
\bibitem{} Kochanek, C. S., \& White, M. 2001, ApJ, 559, 531
\bibitem{} Kochanek, C. S., et al. 2006, ApJ, 640, 47
\bibitem{} Komissarov, S. S., Vlahakis, N. K., Konigl, A., \& Barkov, M. V. 2009, MNRAS in press 
\bibitem{} Kuhlen, M., Keeton, C. R., Madau, P. 2004, ApJ, 601, 104
\bibitem{} Ma, C.-P. 2003, ApJ, 584, L1
\bibitem{} MacFadyen, A. I.\& Woosley, S. E. 1999, ApJ, 524, 262
\bibitem{} Mao, S. 1992, ApJ, 389, L41
\bibitem{} Metcalf, R. B., \& Madau, P. 2001, ApJ, 563, 9
\bibitem{} Nair, S., Jin, C., Garrett, M. A. 2005, MNRAS, 362, 1157
\bibitem{} Nakar, E., Granot, J., Guetta, D. 2004, ApJ, 606L, 37
\bibitem{} Navarro, J., Frank \& White, S. 1996, ApJ, 62, 563
\bibitem{} Nemiroff, R. J., Marani, G. F., Norris, J. P., Bonnell, J. T., Meegan, C. A., \& Hurley, K. C. 2000, in Gamma-Ray Bursts: 5th Huntsville Symposium (AIP Conf. Proc. Vol. 526), p. 663
\bibitem{} Oguri, M. 2002, ApJ, 580, 2
\bibitem{} Oguri, M. \& Keeton, C. R. 2004, ApJ, 610, 663
\bibitem{} Oguri, M. \& Blandford, R. 2009, MNRAS, 392, 930
\bibitem{} Paczynski, B. 1986, ApJ, 308, L43
\bibitem{} Peebles, P. J. E. 1993, ``Principles of physical cosmology'',  
Princeton Series in Physics, Princeton, NJ: Princeton University Press
\bibitem{} Perna, R., Sari, R. \& Frail, D. 2003, ApJ, 594, 379
\bibitem{} Piran, T. 2000, Phys. Rep., 333, 529
\bibitem{} Porciani, C. \& Madau, P. 2001, ApJ, 548, 522 
\bibitem{} Proga, D., Stone, J. M., \& Kallman, T. R. 2000, 2000, ApJ, 613, 686
\bibitem{} Richards, G. T., et al. 2004, ApJ, 610, 679
\bibitem{} Rozo, E., Nagai, D., Keeton, C., \& Kravtsov, A. 2008, ApJ, 687, 22
\bibitem{} Spergel, D. et al. 2007, ApJS, 170, 377
\bibitem{} Takahashi, R., \& Chiba, T. 2001, ApJ, 563, 489
\bibitem{} Warren, M. S., Abazajian, K., Holz, D. E., Teodoro, L. 2006, ApJ, 646, 881
\bibitem{} Winn, J. N., Rusin, D., \& Kochanek, C. S., 2004, Nature, 427, 613

\end{thebibliography}
\end{document}